\documentclass[journal]{IEEEtran}

\ifCLASSINFOpdf
\else
   \usepackage[dvips]{graphicx}
\fi
\usepackage{url}

\usepackage{graphicx, tikz}
\usepackage{amssymb}
\usepackage{xcolor}
\usepackage{mathtools}
\usepackage{hyperref}
\usepackage{csquotes}
\usepackage{amsmath}
\usepackage{amsthm}
\usepackage{utfsym, adforn}
\usepackage{wrapfig}
\usepackage[utf8]{inputenc}
\usepackage{algorithm}
\usepackage{algpseudocode}
\usepackage{graphicx}
\usepackage{subcaption}
\usepackage{booktabs}

\renewcommand{\t}{\mathbf{t}}
\newcommand{\0}{\mathbf{0}}
\newcommand{\x}{\mathbf{x}}
\newcommand{\y}{\mathbf{y}}

\newcommand{\D}{\mathbf{D}}
\newcommand{\U}{\mathbf{U}}
\newcommand{\E}{\mathbf{E}}
\newcommand{\J}{\mathbf{J}}
\newcommand{\I}{\mathbf{I}}
\newcommand{\V}{\mathbf{V}}
\renewcommand{\L}{\mathbf{L}}
\newcommand{\A}{\mathbf{A}}

\renewcommand{\S}{\mathbf{S}}

\renewcommand{\H}{\mathbf{H}}

\newcommand{\w}{\mathbf{w}}
\newcommand{\s}{\mathbf{s}}
\newcommand{\e}{\mathbf{e}}
\renewcommand{\u}{\mathbf{u}}
\newcommand{\bsxi}{\boldsymbol{\xi}}

\newcommand{\z}{\mathbf{z}}

\newcommand{\p}{\mathbf{p}}

\theoremstyle{remark}

\begin{document}

\title{Compensation of correlated autoregressive clock jitter in arrays of Analog-to-Digital Converters}

\author{Daniele Gerosa, Lauri Anttila, and Thomas Eriksson
\thanks{D. G. is a post-doctoral researcher at the Communication, Antennas and Optical Networks, Electrical Engineering dept., Chalmers University of Technology, Göteborg, Sweden (e-mail: daniele.gerosa@chalmers.se).}
\thanks{L. A. is a Senior Research Fellow at Tampere University, Finland (e-mail: lauri.anttila@tuni.fi).}
\thanks{T. E. is full professor at the Communication, Antennas and Optical Networks, Electrical Engineering dept., Chalmers University of Technology, Göteborg, Sweden (e-mail: thomase@chalmers.se).}}

\markboth{}
{Shell \MakeLowercase{\textit{et al.}}: Bare Demo of IEEEtran.cls for IEEE Journals}
\maketitle

\begin{abstract}
In modern communication systems, the fidelity of analog-to-digital converters (ADCs) is limited by sampling clock jitter, i.e., small random timing deviations that undermine ideal sampling. Traditional scalar models often treat jitter as independent Gaussian noise, which makes it essentially untrackable, whereas real ADCs also exhibit temporally correlated (spectrally colored) imperfections. Moreover, spatial cross-correlations between channels in multiple-input multiple-output (MIMO) ADCs are commonly neglected. This paper addresses the joint tracking and compensation of random, cross-correlated timing errors in ADC arrays by modeling jitter as a coupled vector autoregressive process of order one (VAR(1)). We propose a pilot-tone-based Kalman smoother to track and compensate the jitter, and simulations demonstrate substantial reductions in jitter-induced distortion across diverse scenarios.
\end{abstract}

\begin{IEEEkeywords}
MIMO Analog-to-Digital Converters, cross-channel correlated stochastic jitter, vector autoregressive process, Kalman smoother, linearization. 
\end{IEEEkeywords}

\IEEEpeerreviewmaketitle

\section{Introduction}
In modern communication systems, the fidelity of an Analog-to-Digital Converter (ADC) can be fundamentally limited by the stability of its sampling clock. Ideally, samples would be taken precisely at times \( t_n = n T_s \), where \(T_s\) is the nominal sampling interval. In practice, however, each actual sampling instant is offset by a small, random or not, unknown ``clock jitter'' \(\xi_n\), so that \( t_n = n T_s + \xi_n \). The problem of determining whether a sequence \( \{ t_n\}_{n \in \mathbb{N}} \) of sampling points is ``good'' for bandlimited reconstruction has been significantly studied from the abstract perspective of deciding whether the set of perturbed waves \( \{ e^{i t_n \omega} \}_{n \in \mathbb{N}} \) is a frame, or a Riesz basis, for the Hilbert space of square-integrable signals \( L^2 (-\pi,\pi)\). A remarkable result in this direction is a theorem by Kadec \cite{Kadec}, often known as ``Kadec \(1/4\) Theorem'', that states that \( \{ e^{i t_n \omega} \}_{n \in \mathbb{N}} \) is a Riesz basis for \( L^2 (-\pi,\pi) \) as long as \( |n - t_n| \le D < 1/4 \) for all \(n \in \mathbb{N} \) (\(T_s = 1\)). Random perturbations via sequences of i.i.d. random variables have also been considered from a theoretical perspective, see e.g. \cite{Seip_Ulanovskii}. This latter ``model'' is sometimes adopted even in the more applied jitter / phase noise academic literature, where clock jitter deviations are indeed mathematically described with sequences of independent and identically distributed Gaussian random variables \cite{Weller_Goyal}\cite{Testoni_Speciale_Ridolfi_Pouzat}\cite{Araghi_Akhaee_Amini}\cite{Hernandez_Wiesbauer_Paton_Giandomencio}\cite{Kumar_Goyal_Sarma}\cite{Ibarra-Manzanoetal}. Nonetheless, this model fails to fully capture the spectral properties of oscillators as they are measured in testbeds \cite{Zanchi_Samori}. More concretely, ADC clock jitter processes often display piecewise-defined spectra with different slopes: a close-in near DC \( \propto 1/f^3 \) region and a wider \( \propto 1/f^2 \) section that eventually drowns in a flat Gaussian floor; this piecewise model is sometimes referred to as the ``Leeson model'' \cite{Leeson}\cite{Lee_Hajimiri}, even though it was argued in \cite{Piemontese_Colavolpe_Eriksson} that, in communication systems applications, the close-in region could be assumed to be ``flat'' without loss of generality. In sigma-delta converters, clock jitter is shown to have a certain degree of memory that must be accounted for, e.g., via some autoregressive model \cite{Chang_Lin_Wang_Lee_Shih}, a model that appears again in clock phase and frequency synchronization-related works, see \cite{Zino_Dabora_Poor}\cite{Tulone}.

Previous theoretical analyses of \emph{static} jitter in time-interleaved ADCs (Ti ADC) were performed in \cite{elbornsson}\cite{Divi}. The former casts jitter estimation as a heuristic non-convex optimization problem whose solution seeks to restore quasi-stationarity of the signal, while the latter uses Taylor expansions and band-limitation constraints. Cross-correlations between sub-ADCs to estimate static jitter was exploited in \cite{Salib_Flanagan_Cardiff}. The very recent \cite{Sung_Choi} proposes an Extended Kalman Filter-based tracking of Ti ADC time-varying mismatches (including gain imbalances and DC offsets); their formulation however does not account for possible cross-channel correlations, whereas our array-level model will explicitly capture and exploit them, leading to different insights.

In this manuscript we show how sideband pilot-tone injections enable Kalman-smoother-based tracking and removal of clock-jitter-induced distortions in ADC arrays, when the jitter is modeled as a vector autoregressive VAR(1) process with strong cross-channel correlations. The resulting MIMO formulation, which strictly contains per-channel Single-Input Single-Output (SISO) tracking as a special case, exploits the full covariance structure for improved performance and admits a clean analytical treatment that also covers the time-interleaved ADC scenario. 

The system model is introduced and described in Section \ref{sec:SystM} while the algorithm is carefully detailed in Section \ref{sec:AlgDescr}. Section \ref{sec:NumSims} is dedicated to numerically testing our method in diverse scenarios.

\subsection{Notation and symbols.}
We will indicate \emph{column} vectors with bold lowercase letters \( \mathbf{a}, \mathbf{b}, \dots \) and matrices with bold uppercase letters \( \mathbf{A}, \mathbf{B}, \dots \); the \(m\)-th component of the \(n\)-th vector \( \mathbf{a}_n \) will be denoted by \( a_{n,m} \). Positive semi-definiteness for square symmetric matrices will be indicated with the symbol \( \succeq \). The matrix vectorization \( \text{vec}(\mathbf{A}) \) will be obtained by stacking the columns of \( \mathbf{A} \) on top of one another in an \( mn \times 1 \) vector. We will use the symbol \( \otimes \) to denote the Kronecker matrix product, while \( \star \) will be reserved to the convolution between two distributions. \( \delta_{ij} \) will be the Kronecker delta and \( \delta(t) \) will be the Dirac delta distribution.
\section{System model}
\label{sec:SystM}
\subsection{Autoregressive jitter in arrays of Analog-to-Digital Converters}
\label{sec:gen_mod}
\begin{figure}[htbp]
  \centering
\begin{tikzpicture}[line cap=round, line join=round, scale=0.85, thick]

\def\labx{0pt}\def\laby{0pt}

\newcommand{\adc}[4][]{%
  \def\W{1.6}   
  \def\H{0.9}   
  \def\T{0.7}   
  \path
    (#2-1,#3) coordinate (_in)
    (#2+\W,#3) coordinate (_out)
    (#2+0.5*\W,#3) coordinate (_ctr);
  \if\relax\detokenize{#1}\relax\else
    \coordinate (#1-in)     at (_in);
    \coordinate (#1-out)    at (_out);
    \coordinate (#1-center) at (_ctr);
  \fi
  \draw (#2-2,#3) -- (_in);
  \draw (_out) -- (#2+2,#3);
  \draw (#2-1,#3) -- (#2-\T,#3+\H/2) -- (#2+\W,#3+\H/2) --
        (#2+\W,#3-\H/2) -- (#2-\T,#3-\H/2) -- (#2-1,#3);
  \node at (_ctr) [xshift=\labx,yshift=\laby] {$#4$};
}

\def\labx{-10pt}\def\laby{1pt}
\adc[adc1]{0}{1.5}{T_s^{(1)}}

\def\labx{-10pt}\def\laby{1pt}
\adc[adc2]{0}{0}{T_s^{(2)}}

\node at (0.5,-0.65) {$\vdots$};

\def\labx{-10pt}\def\laby{1pt}
\adc[adcM]{0}{-1.5}{T_s^{(M)}}


\node[draw, rectangle, minimum size=10mm, text width=10mm, align=center]
      at (adc2-in) [xshift=-30mm] {shared\\clock};
      
\node at (adc1-in)  [xshift=-15mm] {$x_1(t)$};
\node at (adc1-out) [xshift= 10mm] {$y_{1,n}$};

\node at (adc2-in)  [xshift=-15mm] {$x_2(t)$};
\node at (adc2-out) [xshift= 10mm] {$y_{2,n}$};

\node at (adcM-in)  [xshift=-15mm] {$x_M(t)$};
\node at (adcM-out) [xshift= 10mm] {$y_{M,n}$};

\end{tikzpicture}
\caption{Schematic representation of the system model under consideration}
\label{fig:syst_model}
\end{figure}
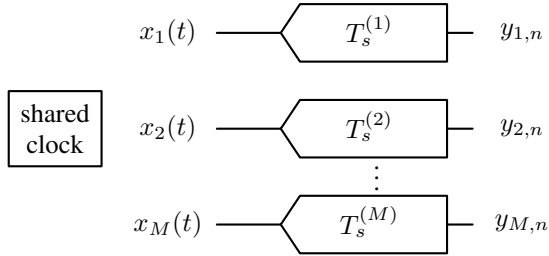

In the present work we extend some of the methods introduced in \cite{Gerosaetal} to parallel arrays of ADCs, where shared clock circuitry introduces correlated autoregressive timing deviations in all the converters. These arrays are schematically represented in Figure \ref{fig:syst_model}; the m-th subADC operates at its nominal sampling frequency \( 1/T_s^{(m)} \), and it samples a waveform \( x_m(t) \) at (ideal) time instants \( n T_s^{(m)}  \). We call \emph{multi-signal} a vector field \( \x : \mathbb{R}^M \to \mathbb{C}^M \) defined by \[ \mathbf{t}_n = (t_{1, n}, \dots , t_{M, n}) \mapsto \x(\t) = (x_1(t_{1, n}), \dots , x_M(t_{M, n})) \]where the \(x_m\) are random bandlimited analog waveforms. We model sampling deviations from the ideal time instants \( (n T_s^{(1)}, \dots , n T_s^{(M)})\) via VAR(1) processes.

 VAR(1) processes generalize AR(1) processes in an ``obvious'' way: let \( \mathbf{V} \in \mathbb{M}_M (\mathbb{R}) \) be an \( M \times M \) real-valued matrix with spectral radius \begin{equation} \label{eq:stab_cond} \rho(\mathbf{V}) = \max_j |\lambda_j (\mathbf{V})| < 1  \end{equation} (stability condition, see \cite{Lutkepohl}); the correlated autoregressive clock jitter process evolves with the law \begin{equation} \label{eq:VAR_def}\boldsymbol{\xi}_n = \mathbf{V}\boldsymbol{\xi}_{n-1} + \boldsymbol{\epsilon}_n  \end{equation} with \( \boldsymbol{\epsilon}_n \sim \mathcal{N}(\boldsymbol{0}, \boldsymbol{\Sigma}_{\boldsymbol{\epsilon}}) \) for all \(n\). The process vector noise can be correlated across the different channels, but it is assumed to be independent in time. 
 
 The spectral density matrix of the process in \eqref{eq:VAR_def} is given by \cite{Lutkepohl} \begin{equation} \label{eq:VAR_psd}
\S(\omega) = \frac{1}{2 \pi }(\I - \V e^{- i \omega})^{-1} \boldsymbol{\Sigma}_{\boldsymbol{\epsilon}} (\I - \V^t e^{ i \omega})^{-1}, \ \omega \in [-\pi, \pi].
 \end{equation}

 The steady-state ``spatial'' covariance matrix \( \boldsymbol{\Xi}_0 = \mathbb{E}[\boldsymbol{\xi}_n \boldsymbol{\xi}_n ^t] =  \text{cov}(\boldsymbol{\xi}_n) \) of the discrete process in \eqref{eq:VAR_def} solves the Lyapunov matrix equation \begin{equation} \label{eq:Lyap} \boldsymbol{\Xi}_0 = \mathbf{V}\boldsymbol{\Xi}_0 \mathbf{V}^t + \boldsymbol{\Sigma}_{\boldsymbol{\epsilon}}; \end{equation} if both \( \boldsymbol{\Sigma}_{\boldsymbol{\epsilon}} \) and \( \mathbf{V} \) are given, an \( \boldsymbol{\Xi}_0 \) that solves \eqref{eq:Lyap} can be found by means of matrix-vectorization, by solving the linear system of equations \( (\mathbf{I} - \mathbf{V} \otimes \mathbf{V}) \text{vec} (\boldsymbol{\Xi}_0) =  \text{vec} (\boldsymbol{\Sigma}_{\boldsymbol{\epsilon}})  \) \cite{Jarlebring}.
 If instead a \( \boldsymbol{\Xi}_0 \succ 0 \) is measured or accurately estimated, there are infinitely many pairs \( (\V, \boldsymbol{\Sigma}_\epsilon) \) that solves \eqref{eq:Lyap}, because the cone of (strictly) positive definite matrices is open in any matrix norm topology, and thus if \( \| \V \|_F \) is small enough then \(\boldsymbol{\Xi}_0 - \V \boldsymbol{\Xi}_0 \V^t \succeq 0 \), which is also symmetric because \( \boldsymbol{\Xi}_0 \) is so. 

  The situation is different if both \( \boldsymbol{\Xi}_0 \) \emph{and} \(\boldsymbol{\Xi}_1 = \mathbb{E}[\boldsymbol{\xi}_n \boldsymbol{\xi}_{n-1} ^t ] = \V \boldsymbol{\Xi}_0  \) are measured or in general available; in that case, assuming again \( \boldsymbol{\Xi}_0 \succ 0 \), we have \( \V = \boldsymbol{\Xi}_1 \boldsymbol{\Xi}_0 ^{-1} \) and \( \boldsymbol{\Sigma}_\epsilon = \boldsymbol{\Xi}_0 - \boldsymbol{\Xi}_1 \boldsymbol{\Xi}_0 ^{-1} \boldsymbol{\Xi}_1 ^t \) from \eqref{eq:Lyap}. These can be seen as Yule-Walker equations for vector processes and allow process parameters estimation from their covariance properties in low noise scenarios.

\subsection{Example: the VAR model for Time-interleaved Analog-to-Digital Converters}
Time-interleaved Analog-to-Digital Converters raise the effective sampling rate by operating many converter channels in parallel, each observing the same signal at staggered time instants. Practical implementations suffer from gain, offset, and bandwidth mismatches, as well as timing errors (static skew and random jitter), which create spurious tones and reduce resolution \cite{elbornsson}\cite{Divi}. Because clocking is typically shared, timing deviations are often correlated across channels, making accurate modeling essential. Here we adopt a two-step view: ideal interleaving as pure analog delays of a reference path, followed by sampling with small random timing perturbations. 

The general abstract model outlined in Section \ref{sec:gen_mod} can be particularized to describe the dynamics of a Ti ADC, in a way in which we now illustrate. First, the interleaving action on the analog stimulus can be represented as convolution between the reference channel and an appropriate analog delay filter \( h_m \) (e.g. \( h_m(t) = \delta(t + (m-1)T_s)\)), as described in \cite{Papoulis}; namely \[ x_m(t) = x_1(t + (m-1)T_s) = (x_1 \star h_m)(t) \]    for \( m =2, \dots , M  \). \( T_s \) is the ``fast'' sampling time so that \( T_s ^{(m)} = M T_s \) for all \(m\); the sampling vector instants will then simply be \( \t_n = (nMT_s, \dots ,nMT_s)  \). 

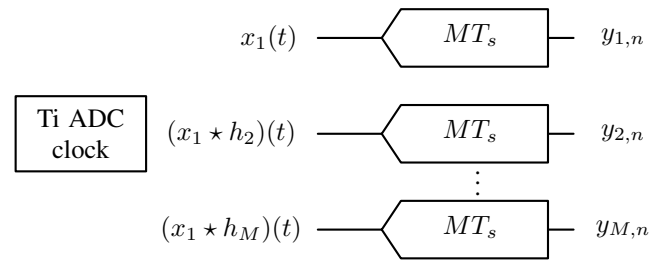
\begin{figure}[htbp]
  \centering
\begin{tikzpicture}[line cap=round, line join=round, scale=0.85, thick]

\def\labx{0pt}\def\laby{0pt}

\newcommand{\adc}[4][]{%
  \def\W{1.6}   
  \def\H{0.9}   
  \def\T{0.7}   
  \path
    (#2-1,#3) coordinate (_in)
    (#2+\W,#3) coordinate (_out)
    (#2+0.5*\W,#3) coordinate (_ctr);
  \if\relax\detokenize{#1}\relax\else
    \coordinate (#1-in)     at (_in);
    \coordinate (#1-out)    at (_out);
    \coordinate (#1-center) at (_ctr);
  \fi
  \draw (#2-2,#3) -- (_in);
  \draw (_out) -- (#2+2,#3);
  \draw (#2-1,#3) -- (#2-\T,#3+\H/2) -- (#2+\W,#3+\H/2) --
        (#2+\W,#3-\H/2) -- (#2-\T,#3-\H/2) -- (#2-1,#3);
  \node at (_ctr) [xshift=\labx,yshift=\laby] {$#4$};
}

\def\labx{-10pt}\def\laby{1pt}
\adc[adc1]{0}{1.5}{M T_s}

\def\labx{-10pt}\def\laby{1pt}
\adc[adc2]{0}{0}{M T_s}

\node at (0.5,-0.65) {$\vdots$};

\def\labx{-10pt}\def\laby{1pt}
\adc[adcM]{0}{-1.5}{M T_s}


\node[draw, rectangle, minimum size=10mm, text width=15mm, align=center]
      at (adc2-in) [xshift=-40mm] {Ti ADC\\clock};
      
\node at (adc1-in)  [xshift=-15mm] {$x_1(t)$};
\node at (adc1-out) [xshift= 10mm] {$y_{1,n}$};

\node at (adc2-in)  [xshift=-20mm] {$(x_1 \star h_2)(t)$};
\node at (adc2-out) [xshift= 10mm] {$y_{2,n}$};

\node at (adcM-in)  [xshift=-20mm] {$(x_1 \star h_M)(t)$};
\node at (adcM-out) [xshift= 10mm] {$y_{M,n}$};

\end{tikzpicture}
\caption{Schematic representation of a Time-interleaved Analog-to-Digital Converter.}
\label{fig:ti_adc}
\end{figure}

Ti ADCs are typically regulated by a single clock, leading to the simplest scenario in which jitter disturbances collapse into a single process whose samples are picked sequentially by the \(M\) interleaved channels; consequently the m-th sub-ADC inherits the previous sub-ADC's jitter within the same frame and the chain wraps across frames. Mathematically expressed we simply have \begin{equation} \label{eq:jitt_ti} \xi_{m,n} = \begin{cases} \varphi \xi_{m-1,n} + \epsilon_{m,n} & \text{if } m \ge 2, \\ \varphi \xi_{M,n-1} + \epsilon_{1,n} & \text{if } m= 1 \end{cases}  \end{equation} where \( \boldsymbol{\epsilon}_n = (\epsilon_{1,n}, \dots , \epsilon_{M,n})^t \sim \mathcal{N}(\mathbf{0}, \boldsymbol{\Sigma}_{\boldsymbol{\epsilon}}) \) and \( \varphi \in (0,1) \).
The challenge, that we now address, is to express \eqref{eq:jitt_ti} compactly with the VAR formalism \eqref{eq:VAR_def}. Let \( \E \) be the matrix defined by \( \E_{1M} = 1 \) and zero elsewhere; let moreover \( \J \) be the matrix defined by \( \J_{1m} = 0 \) for all \( m = 1, \dots, M \) and \( \J_{jm} = \delta_{(j-1)m}  \) for \( j \ge 2 \), \( m \ge 1 \). Then the sought Ti ADC jitter dynamics can be expressed as \begin{equation} \begin{split}
\boldsymbol{\xi}_n & = \varphi \J \boldsymbol{\xi}_n + \varphi \E \boldsymbol{\xi}_{n-1} + \boldsymbol{\epsilon}_n \\    & = (\I - \varphi \J)^{-1} \varphi\E \boldsymbol{\xi}_{n-1} + (\I - \varphi \J)^{-1}\boldsymbol{\epsilon}_n.
\end{split}
\end{equation}
The matrix \( \J \) is nilpotent of index \(M\), meaning that \( \J^M = \0 \) but \( \J^m \ne \0 \) if \( m < M \); therefore \( (\I - \varphi \J)^{-1} \) can be explicitly computed using the Neumann series \cite{Horn_Johnson} \[ (\I - \varphi \J)^{-1} = \sum_{m=0}^{M-1} \varphi^m \J^m.  \] The latter formula can be used to derive the product \( (\I - \varphi \J)^{-1} \E  \); indeed \( \J^m \E =  \e_{m+1} \e_M ^t \) leading to \[ (\I - \varphi \J)^{-1} \E = \sum_{m=0}^{M-1}\varphi^m \e_{m+1} \e_M ^t = \u  \e_M ^t \] with \( \u = (1, \varphi, \dots , \varphi^{M-1})^t \). \( \V =  \varphi (\I - \varphi \J)^{-1} \E \) has rank \(1\) and \( \rho(\V) = \varphi^M \), which fulfills the spectral stability condition \eqref{eq:stab_cond}. Moreover the term \( \boldsymbol{\eta}_n = (\I - \varphi \J)^{-1} \boldsymbol{\epsilon}_n \) is Gaussian with covariance matrix \(  (\I - \varphi \J)^{-1} \boldsymbol{\Sigma}_{\boldsymbol{\epsilon}} ((\I - \varphi \J)^t)^{-1}  \).

The conclusion is that the jitter process in Ti ADCs can be compactly and conveniently rewritten using the VAR(1) formalism introduced earlier \begin{equation} \label{eq:var_proc_tiadc} \begin{split}
\boldsymbol{\xi}_n & = \varphi \u \e_M ^t \boldsymbol{\xi}_{n-1} + \boldsymbol{\eta}_n \\ & = \varphi \u  \xi_{M,n-1} + \boldsymbol{\eta}_n.
\end{split}
\end{equation}

\subsubsection{Steady-state covariance matrix}
The steady-state covariance matrix of the Ti ADC jitter model built in \eqref{eq:var_proc_tiadc} can be easily derived as solution to Equation \eqref{eq:Lyap} by setting \( \V = \varphi \u \e_M ^t \) and \( \boldsymbol{\Sigma}_{\boldsymbol{\eta}} =  (\I - \varphi \J)^{-1} \boldsymbol{\Sigma}_{\boldsymbol{\epsilon}} ((\I - \varphi \J)^t)^{-1}  \); \eqref{eq:Lyap} becomes then \[  \boldsymbol{\Xi}_0 - \varphi^2(\boldsymbol{\Xi}_0)_{MM} \u \u^t = \boldsymbol{\Sigma}_{\boldsymbol{\eta}} \] from which it must be \( (\boldsymbol{\Xi}_0)_{MM} = (\boldsymbol{\Sigma}_{\boldsymbol{\eta}})_{MM} / (1 - \varphi^{2M}) \). Thus \begin{equation} \label{eq:tiadc_steadystatecov} \boldsymbol{\Xi}_0 = \frac{\varphi^2 (\boldsymbol{\Sigma}_{\boldsymbol{\eta}})_{MM}}{1 - \varphi^{2M}} \u \u^t + \boldsymbol{\Sigma}_{\boldsymbol{\eta}}. \end{equation}

\section{Pilot tones based tracking and compensation algorithm}
\label{sec:AlgDescr}
The impact of vector jitter \eqref{eq:VAR_def} on a multi-signal can be assessed in the same fashion as done in the 1-D case in \cite{Gerosaetal}, by means of a first-order Taylor expansion: \begin{equation} \label{eq:first_ord}
    \x(\t_n + \bsxi_n) \approx \x(\t_n) + \D_{\x} (\t_n) \bsxi _n
\end{equation}
where \(\D_{\x}\) is the (diagonal) Jacobian matrix of \( \x \), i.e. \(\D_{\x} = \text{diag}(x_1 '(t), \dots, x_M ' (t)) \), and \( \t_n = (nT_s^{(1)}, \dots , n T_s ^{(M)}) \). Attempts to go beyond the first-order term have been made at least in \cite{Divi_Wornell}\cite{Shenetal}, even though the overhead of cascading multiple first-order differentiating filters could be significant and introduce latency in the system. 

Our sampled vector signal model also contains additive white noise, and thus the model adopted here and henceforth is \begin{equation} \label{eq:sig_model}
\y_n = \x(\t_n) + \D_{\x} (\t_n) \bsxi _n + \w_n,
\end{equation} \(\w_n \sim \mathcal{N}(\0, \boldsymbol{\Sigma}_\w) \),  \(n \in \mathbb{N} \).

The estimation and compensation algorithm that we propose to de-jitter a multi-signal is based on the injection of one out-of-band high frequency pilot tone per-channel, in a similar fashion as done in e.g. \cite{Towfic_Ting_Sayed}\cite{Syrjälä_Valkama} in the single channel scenario. The novelty of the proposed approach lies particularly in the jitter dynamics modeling, which allows us to cast the process tracking problem in a control-theoretical sense that accounts for cross-channel correlations. More concretely let \( s_m(t) \) be the unknown baseband analog payloads bandlimited to \([-W_m ^{(p)},W_m ^{(p)}]\) and \( p_m(t) = A_m e^{2 \pi i f_{0,m} t} \) single pilot tones such that \( f_{0, m} \gg W_m ^{(p)} \). We assume that the received and sampled multi-signal is, following \eqref{eq:sig_model}, \begin{equation} \label{eq:final_model} \y_n = \underbrace{\s(\t_n) + \p(\t_n)}_{= \x(\t_n)} + \D_{\s + \p} (\t_n) \boldsymbol{\xi}_n + \w_n. \end{equation} Notice that \( \D_{\s + \p} (\t_n) = (\D_{\s} + \D_{ \p}) (\t_n) \). Both the payload and the pilot tone are affected by the same jitter process, and therefore the idea is to utilize the full knowledge of \( \p \) to obtain an estimation of \( \boldsymbol{\xi}_n \); the pilots must be isolated from the payloads with a band-pass filter, whose bandwidth \( [f_{0,m} - W^{(bp)}_m, f_{0,m} + W^{(bp)} _m] \) must be large enough to ``let pass'' enough jitter power for the estimation process to effectively take place (without taking in portions of the payload though): indeed the power spectral density of \( s_{m,n}' \xi_{m,n}  \) is far from being concentrated in the pilot's bin. If we indicate with \( \boldsymbol{\mathcal{H}} \) an ideal band-pass filter that bandpasses each channel of \( \y \), we will have \[ \begin{split} \z_n =  \boldsymbol{\mathcal{H}}(\y _n) & = \p(\t_n) + \boldsymbol{\mathcal{H}}(\D_{\p} (\t_n) \boldsymbol{\xi}_n + \w_n) \\ & \approx \p(\t_n) + \D_{\p} (\t_n) \boldsymbol{\xi}_n + \boldsymbol{\mathcal{H}}(\w_n). \end{split}  \]

With this at hand we can finally cast the state-space representation of the problem under consideration: 

\begin{equation}
\label{eq:st_sp_VAR}
\begin{cases}
\boldsymbol{\xi}_n & =\V \boldsymbol{\xi}_{n-1} + \boldsymbol{\epsilon}_n \\
\z_n & = \p(\t_n) + \D_\p (\t_n) \boldsymbol{\xi}_n + \boldsymbol{\mathcal{H}}(\w_n);
\end{cases}
\end{equation}
notice that \( \D_\p (\t_n) = 2\pi i \text{ diag}( \{A_m f_{0,m} e^{2 \pi i f_{0,m} n T_s^{(m)}} \}_{m=1} ^M)  \). 

Once a valid estimate \( \widehat{\boldsymbol{\xi}}_n \) of the (realization of) the process \(\boldsymbol{\xi}_n  \) is obtained, for example by means of a Kalman smoother that we use here in the Rauch-Tung-Striebel (RTS) form,  we de-jitter the received multi-signal \eqref{eq:final_model} in two steps: first by removing the jittered pilot tone \[ \overline{\y}_n  =  \y_n - (\p(\t_n) + \D_p(\t_n)\widehat{\boldsymbol{\xi}}_n)  \] and subsequently by removing the distortion caused by the jitter in the payload itself \begin{equation} \label{eq:out} \overline{\overline{\y}}_n = \overline{\y}_n - \D_s (\t_n) \widehat{\boldsymbol{\xi}}_n;   \end{equation} we stress here that \( \D_\s (\t_n) \) will in general be unknown. 

\begin{algorithm}[H]
\caption{Kalman filter and RTS smoother for \eqref{eq:st_sp_VAR}}
\label{alg:tiadc_kalman_rts}
\begin{algorithmic}[1]
\Require $\V\!\in\!\mathbb{R}^{M\times M}$, $\boldsymbol{\Sigma}_{\boldsymbol{\epsilon}} \succeq \0 $, $\boldsymbol{\Sigma}_{\boldsymbol{\mathcal{H}}(\w_n)}\succeq \0$; sequences $\{\p(\t_n)\}_{n=1}^N$, $\{\z_n\}_{n=1}^N$, and $\{\D_\p(\t_n)\}_{n=1}^N$
\State \textbf{Initialization (steady-state prior):} find $\mathbf{P}_{0|0}$ solving $\mathbf{P}_{0|0}=\V \mathbf{P}_{0|0}\V^{t}+\boldsymbol{\Sigma}_{\boldsymbol{\epsilon}}$; set $\boldsymbol{\mu}_{0|0}=\mathbf{0}$
\State $\mathbf{I}_M \gets \text{identity in }\mathbb{R}^{M\times M}$
\For{$n=1,\dots,N$}
  \State $\boldsymbol{\mu}_{n|n-1}=\V\,\boldsymbol{\mu}_{n-1|n-1}$ \label{line:pred_mu}
  \State $\mathbf{P}_{n|n-1}=\V\,\mathbf{P}_{n-1|n-1}\,\V^{t}+\boldsymbol{\Sigma}_{\boldsymbol{\epsilon}}$ \label{line:pred_P}
  \State $\mathbf{H}_n \gets \D_\p(\t_n)$ 
  \State $\widetilde{\mathbf{z}}_n \gets \z_n-\p(\t_n)$ 
  \State $\boldsymbol{\nu}_n=\widetilde{\mathbf{z}}_n-\mathbf{H}_n\,\boldsymbol{\mu}_{n|n-1}$ \Comment{innovation}
  \State $\mathbf{S}_n=\mathbf{H}_n\mathbf{P}_{n|n-1}\mathbf{H}_n^{t}+\boldsymbol{\Sigma}_{\boldsymbol{\mathcal{H}}(\w_n)}$ \Comment{innovation covariance}
  \State $\mathbf{K}_n=\mathbf{P}_{n|n-1}\mathbf{H}_n^{t}\mathbf{S}_n^{-1}$ \Comment{Kalman gain}
  \State $\boldsymbol{\mu}_{n|n}=\boldsymbol{\mu}_{n|n-1}+\mathbf{K}_n\boldsymbol{\nu}_n$
  \State $\mathbf{P}_{n|n}=(\mathbf{I}_M-\mathbf{K}_n\mathbf{H}_n)\mathbf{P}_{n|n-1}(\mathbf{I}_M-\mathbf{K}_n\mathbf{H}_n)^{t}+\mathbf{K}_n\boldsymbol{\Sigma}_{\boldsymbol{\mathcal{H}}(\w_n)}\mathbf{K}_n^{t}$ 
\EndFor
\State \textbf{Filtered outputs:} $\widehat{\boldsymbol{\xi}}^{\text{filt}}_n\gets\boldsymbol{\mu}_{n|n}$, \ $\mathbf{P}^{\text{filt}}_n\gets\mathbf{P}_{n|n}$ \quad for $n=1,\dots,N$
\Statex
\State \textbf{RTS smoother (backward pass):}
\State $\widehat{\boldsymbol{\xi}}^{\text{smooth}}_N=\widehat{\boldsymbol{\xi}}^{\text{filt}}_{N}$, \ $\mathbf{P}^{\text{smooth}}_N=\mathbf{P}^{\text{filt}}_{N}$
\For{$n=N-1,\dots,1$}
  \State $\mathbf{J}_n=\mathbf{P}_{n|n}\V^{t}\mathbf{P}_{n+1|n}^{-1}$ \Comment{smoother gain}
  \State $\widehat{\boldsymbol{\xi}}^{\text{smooth}}_{n}=\widehat{\boldsymbol{\xi}}^{\text{filt}}_{n}+\mathbf{J}_n\!\left(\hat{\boldsymbol{\xi}}^{\text{smooth}}_{n+1}-\boldsymbol{\mu}_{n+1|n}\right)$
  \State $\mathbf{P}^{\text{smooth}}_{n}=\mathbf{P}_{n|n}+\mathbf{J}_n\!\left(\mathbf{P}^{\text{smooth}}_{n+1}-\mathbf{P}_{n+1|n}\right)\mathbf{J}_n^{t}$
\EndFor
\State \textbf{Return:} $\{\widehat{\boldsymbol{\xi}}_n = \widehat{\boldsymbol{\xi}}^{\text{smooth}}_n\}_{n=1}^N$
\end{algorithmic}
\end{algorithm}

Therefore we replace it with \( \D_{\overline{\y}} (\t_n) \), that can be computed by passing \( \overline{\y} \) through an (ideal) differentiator filter. More advanced differentiating filters, e.g. Savitzky-Golay, might offer superior smoothing properties but they are not employed in the present. Whether \( \D_\s (\t_n) \approx \D_{\overline{\y}} (\t_n) \) holds will depend in general on \( \boldsymbol{\Sigma}_\w \) and on \( \boldsymbol{\Xi}_0  \); however for the first-order expansion in \eqref{eq:first_ord} to hold confidently, the per-channel jitter cannot ``heuristically'' exceed a fraction of the sampling interval, namely \( \sqrt{(\boldsymbol{\Xi}_0)_{mm}} / T_s ^{(m)} \ll 1 \). If the latter holds, then it is possible to replace \( \D_\s (\t_n) \) with \( \D_{\overline{\y}} (\t_n) \), as rigorously shown in Section II.B of \cite{Gerosaetal}.

We now summarize the proposed algorithm in the following chart:
\begin{algorithm}[H]
\caption{Pilot tones-aided algorithm for VAR jitter estimation and compensation}
\label{alg:full_alg}
\begin{algorithmic}[1]
\Require $\V\!\in\!\mathbb{R}^{M\times M}$, $\boldsymbol{\Sigma}_{\boldsymbol{\epsilon}} \succeq \0$, $\boldsymbol{\Sigma}_{\boldsymbol{\mathcal{H}}(\w_n)}\succeq \0$; sequences $\{\p(\t_n)\}_{n=1}^N$, $\{\z_n\}_{n=1}^N$, and $\{\D_\p(\t_n)\}_{n=1}^N$; band-pass filter \(\boldsymbol{\mathcal{H}} \)
\State Apply the filter \( \boldsymbol{\mathcal{H}} \) to \( \y_n \);
\State Use Algorithm \ref{alg:tiadc_kalman_rts} to extract an estimate \( \widehat{\boldsymbol{\xi}}_n \) of \(\boldsymbol{\xi}_n  \) from \( \z_n \);
\State Use \( \widehat{\boldsymbol{\xi}}_n \) to de-jitter the pilot tones, and get \(\overline{\y}_n \);
\State Compute \( \D_{\overline{\y}} \) and use it to remove the jitter distortion from the payload;
\State \textbf{Return:} \( \overline{\overline{\y}}_n \) \eqref{eq:out}

\end{algorithmic}
\end{algorithm}

\section{Numerical Simulations}
\label{sec:NumSims}
In this section we validate the proposed model and algorithm with some numerical simulations. In a baseline experiment we generated \( M=8 \) OFDM waveforms with \(N= 2^{16}\) samples and, per channel, a random number of active carriers ranging from \(13 000\) to \(14 000\); the data symbols were drawn uniformly at random from a \(16\) QAM constellation. We set \( f_s^{(m)} = 100 \text{ MSPs} \) for all \(m\) and we superimposed the pilot tones to the unknown payload at \( 30 \text{ MHz} \) in each channel. As target performance metrics we used the average Signal-to-Jitter-Distortion Ratio (SJDR) pre- and post-compensation: 

\begin{equation} \label{eq:SJDR_pre}
    \mathrm{SJDR}_\mathrm{pre} = \frac{10}{M} \sum_{m=1}^M  \log_{10} \left[ \frac{ \frac{1}{N} \sum_{n=1}^N s_{n,m}^2 }{\frac{1}{N} \sum_{n=1}^N(s' _{n,m}  \xi_{n,m})^2 }   \right] 
\end{equation}
and 
\begin{equation} \label{eq:SJDR_post}
    \mathrm{SJDR}_\mathrm{post} =\frac{10}{M} \sum_{m=1}^M  \log_{10} \left[ \frac{ \frac{1}{N} \sum_{n=1}^N s_{n,m}^2 }{\frac{1}{N} \sum_{n=1}^N(s' _{n,m}  \xi_{n,m} - y' _{n,m} \widehat{\xi}_{n,m})^2 }   \right] 
\end{equation}
and analogously the Signal-to-Noise and Distortion Ratio (SINADR) where also the white noise powers are included in the denominators of \eqref{eq:SJDR_pre} and \eqref{eq:SJDR_post}. All these metrics were additionally averaged in a Monte Carlo simulation with \(100\) different noise realizations per channel. 

The vector jitter dynamics \eqref{eq:VAR_def} representing a generic MIMO ADC was generated so that \( \boldsymbol{\Xi}_0 \) is a dense matrix with \emph{prescribed} diagonals (that set the average amount of jitter per-channel so that it does not exceed a certain percentage of the sampling interval) and high cross-correlation between all \(M\) channels. This was obtained by fixing (i.e. deciding a priori and then keeping it fixed throughout the whole Monte Carlo simulation) the matrix \( \boldsymbol{\Xi}_0 \); the matrix \(\V  \) was set equal to \( \L \U \A \U ^t \L^{-1} \) with \( \A \) diagonal containing numbers close to \(1\), \(\U\) random unitary and \( \L \) coming from the Cholesky factorization of \( \boldsymbol{\Xi}_0 \).

\begin{figure}[ht!]
  \centering
  \includegraphics[width=0.49\textwidth]{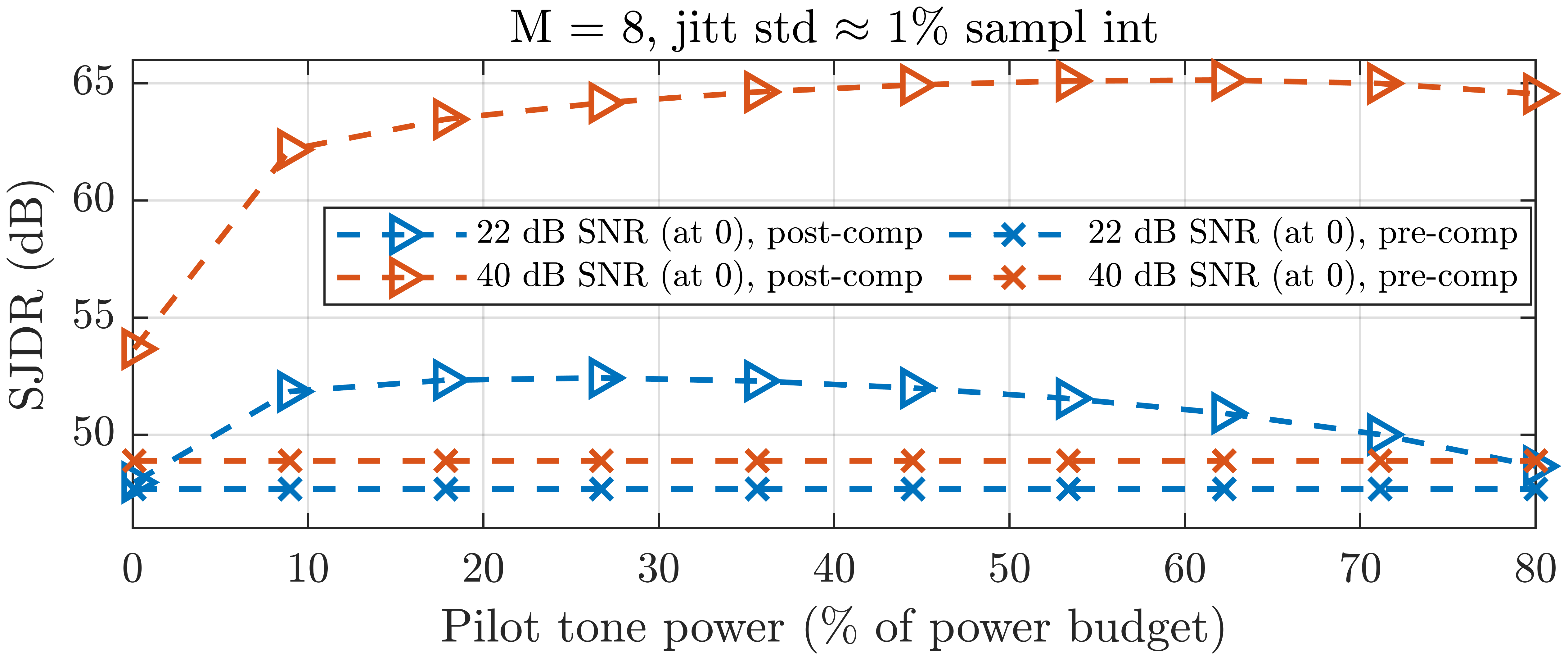}\\[1.2em]
  \includegraphics[width=0.49\textwidth]{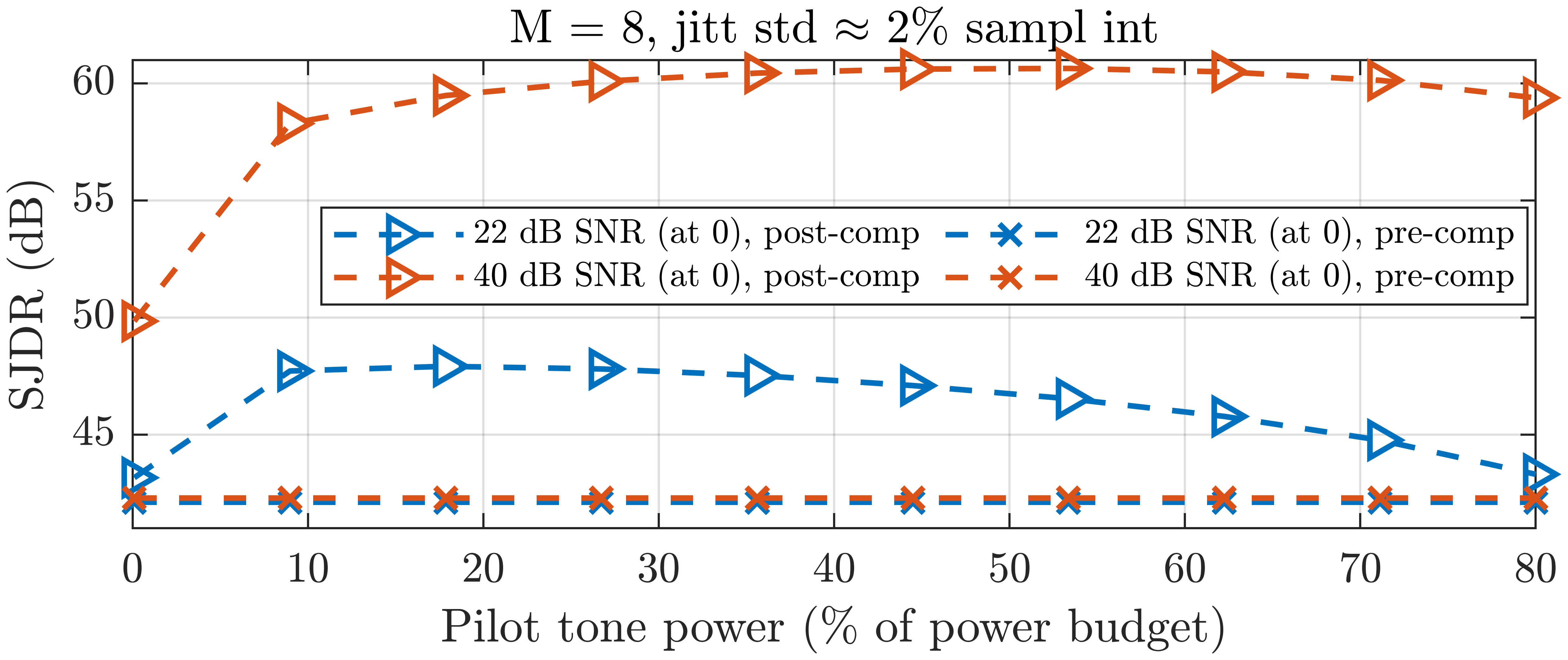}
  \caption{Low jitter, high noise. The leftmost power percentage is \(0.999\).}
  \label{fig:pow_budg_SJDR}
\end{figure}

Then \( \boldsymbol{\Sigma}_{\boldsymbol{\epsilon}} = \boldsymbol{\Xi}_0 - \V \boldsymbol{\Xi}_0 \V^t = \L (\I - \U \A \A^t \U^t) \L^t \), and the latter is symmetric and positive semi-definite because so is \( (\I - \U \A \A^t \U^t) \) and conjugacy preserves both properties. Throughout this whole section the method is assumed to have complete knowledge of \( \V \), \( \boldsymbol{\Sigma}_{\boldsymbol{\epsilon}} \) and \( \boldsymbol{\Sigma}_{\w} \)

\begin{figure}[ht!]
  \centering
  \includegraphics[width=0.49\textwidth]{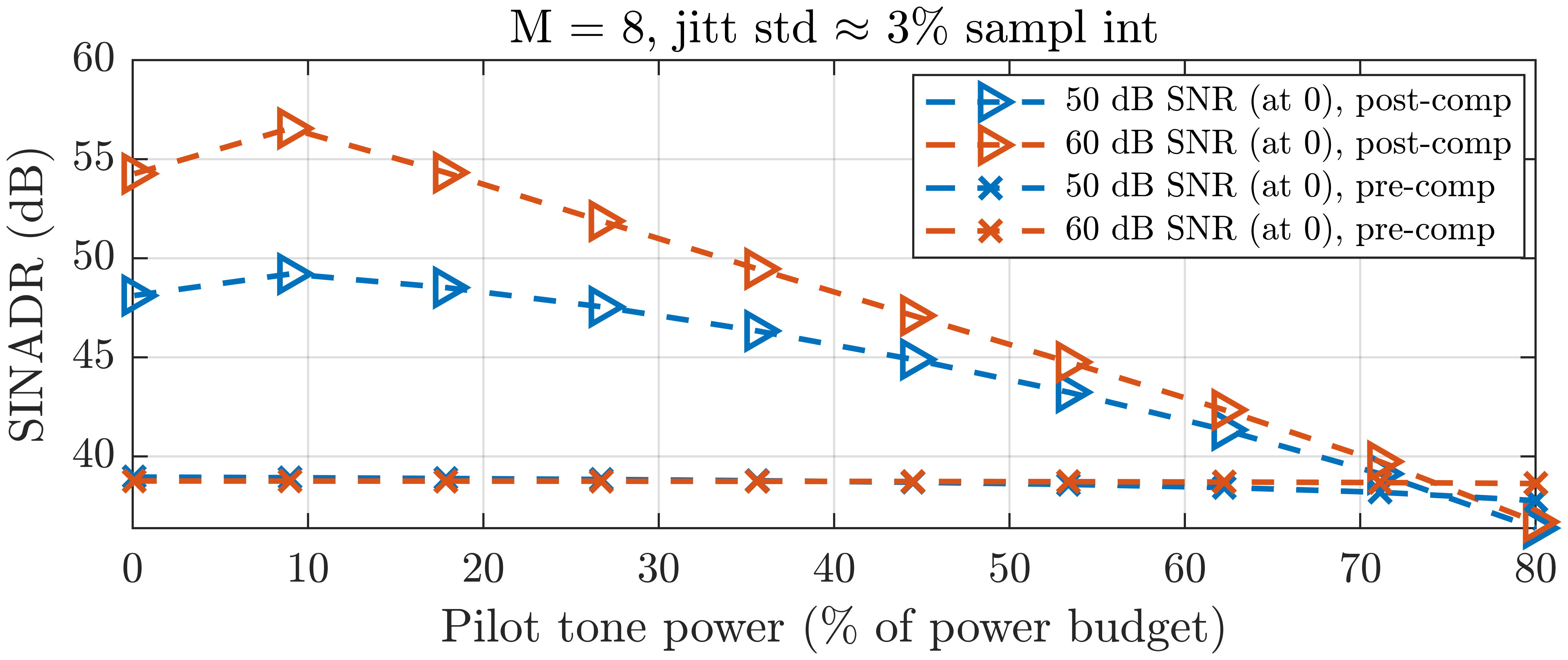}\\[1.2em]
  \includegraphics[width=0.49\textwidth]{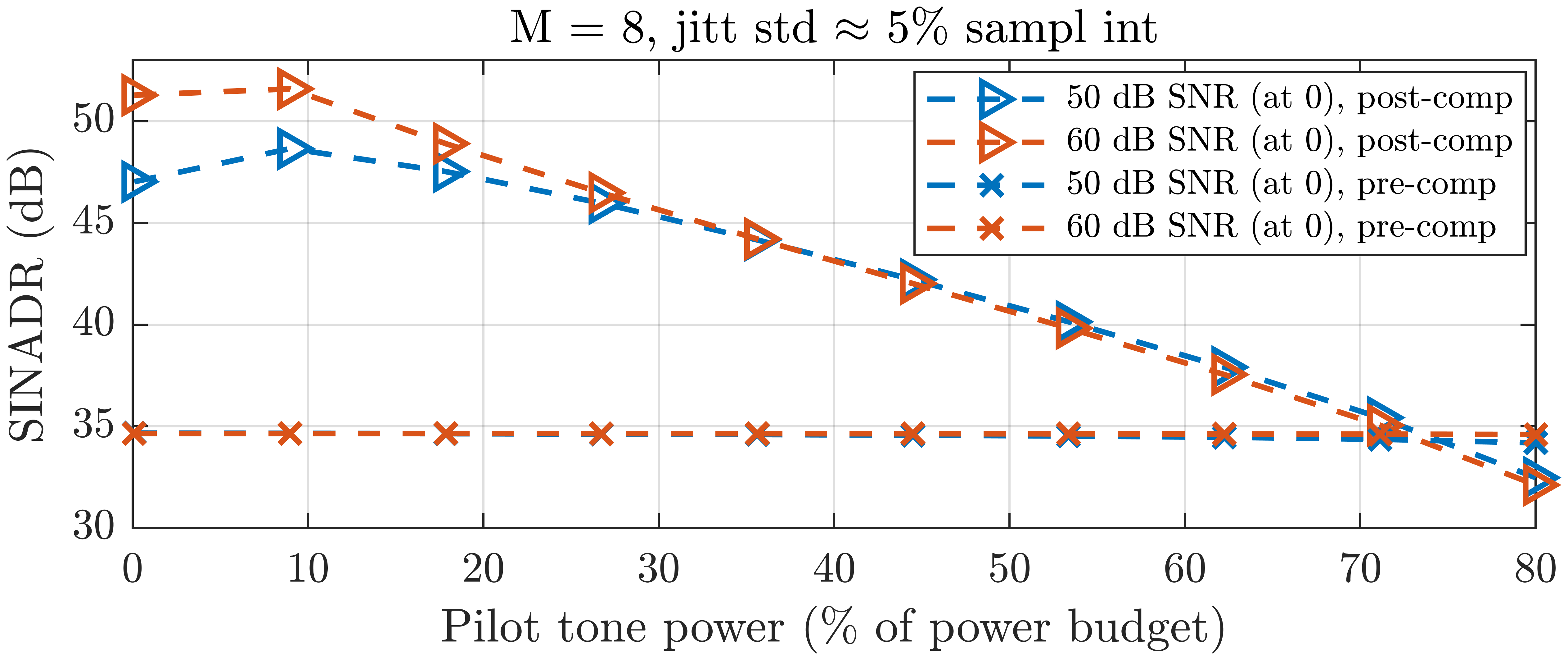}
  \caption{High jitter, low noise. The leftmost power percentage is \(0.999\).}
  \label{fig:pow_budg_SINADR}
\end{figure}

The main goal was here to assess the behavior of the algorithm under a power constraint, meaning that we assumed a (unitary) power budget to be shared between payloads and pilots. Figure \ref{fig:pow_budg_SJDR} summarizes this simulation. Two noise levels are considered here, corresponding - when no pilots are present - to SNR \(22\) and \(40\) dB respectively, via a constant-power additive white noise. The algorithm performances begin to decrease after a certain pilot power level since the SNR progressively decreases and with it the fidelity of the approximation \( \D_\s \approx \D_{\y'} \), that becomes invalid. 
Figure \ref{fig:pow_budg_SJDR} shows that the proposed algorithm tracks and suppresses autoregressive jitter even under substantial white-noise levels. In these noise-dominated settings, the gain in SINADR is negligible because the noise power exceeds the jitter-distortion power across all cases. To make SINADR effects observable, we therefore repeated the sweep in a different regime by setting the additive-noise SNR very high and increasing the jitter variance up to \( \approx 5 \%\) of the sampling interval (with the risk of ``breaking'' the linearization model in \eqref{eq:first_ord}), and the results are showcased in Figure \ref{fig:pow_budg_SINADR} from where we deduce that, in the scenarios evaluated, the optimal power level allocation for pilots must not exceed \(10 \%\) of the total available power. 

\begin{figure}[ht!]
  \centering
  \includegraphics[width=0.49\textwidth]{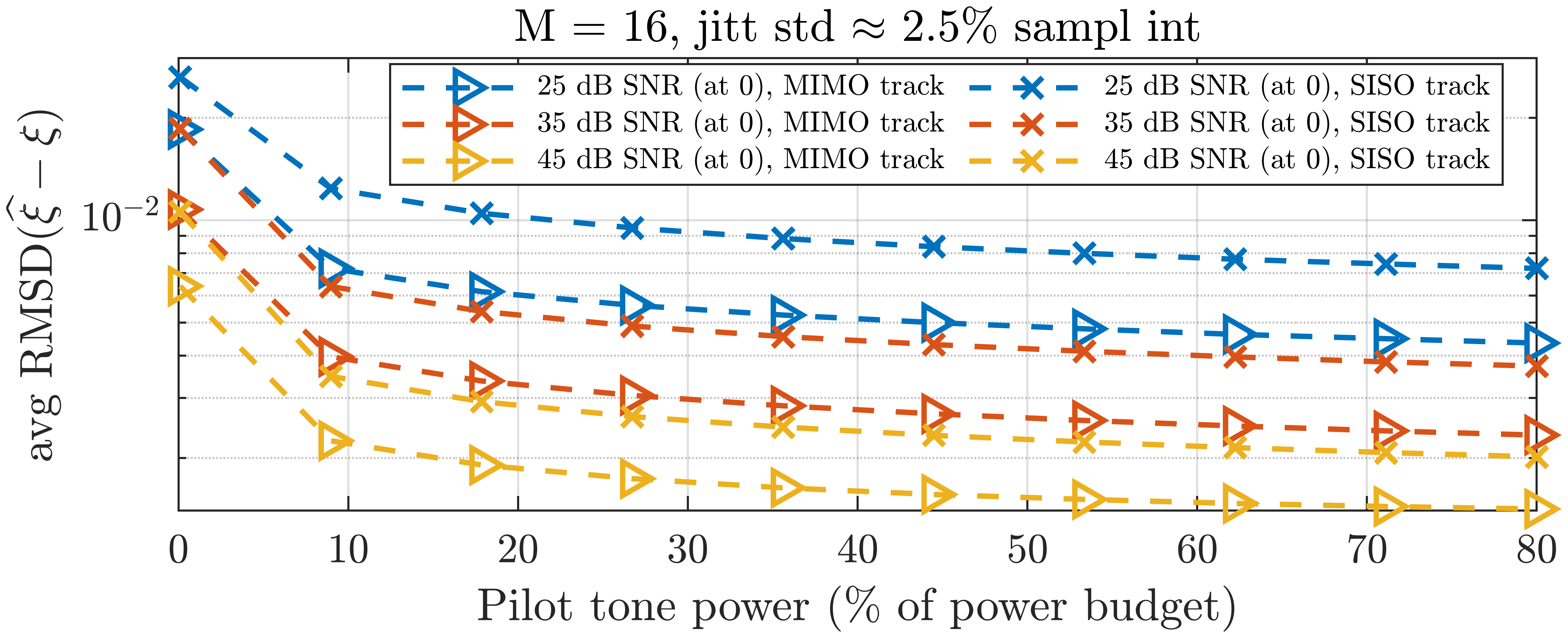}
  \caption{MIMO vs SISO tracking.}
  \label{fig:pow_budg_MIMOvsSISO}
\end{figure}

In Figure \ref{fig:pow_budg_MIMOvsSISO} we instead plot the average (across the \(M\) channels) Root Mean Square Deviation (RMSD) of the estimates \( \widehat{\bsxi}  \) with respect to the ground truth process \begin{equation}
    \label{eq:avgRMSD}
    \text{avg RMSD}(\widehat{\bsxi} - \bsxi) = \frac{1}{M} \sum_{m = 1}^M \sqrt{ \frac{1}{N} \sum_{n=1}^N (\widehat{\xi}_{n,m} - \xi_{n,m})^2  }
\end{equation}
in three different SNR scenarios, and we again sweep through different power level combinations. The SISO “track” in the legend refers to per-channel tracking that does not exploit cross-correlation information, i.e., we perform \(M\) independent 1-D tracks. As expected, exploiting the second-order statistics of the hidden process is fundamental for optimal estimation and directly reflects in overall jitter reduction. In particular, the “MIMO” curves systematically dominate the “SISO” ones, confirming that exploiting the off-diagonal structure of \( \boldsymbol{\Xi}_0 \) in the Kalman smoother yields a clear tracking gain whenever the jitter processes are correlated across channels. Intuitively, the joint model lets each channel “borrow strength” from the others, improving jitter tracking accuracy without extra pilot power or a higher-order temporal model.

\section{Conclusions}
In this paper we addressed de-jittering of MIMO ADC arrays by modeling receiver timing errors as a cross-correlated VAR(1) process, \(\bsxi_n=\V\bsxi_{n-1}+\boldsymbol{\epsilon}_n\), which captures spectrally colored spectra. We proposed a pilot-tone aided Kalman smoother to estimate \(\widehat{\bsxi}_n\) from \(\z_n \approx \p(\t_n)+\D_{\p}(\t_n)\bsxi_n+\mathcal{H}(\w_n)\) and a payload de-jittering scheme. We showed that, in noise-limited conditions, SJDR increases while SINADR remains essentially constant, whereas in jitter-limited conditions both improve markedly. Exploiting off-diagonal structure in \(\boldsymbol{\Xi}_0\) (MIMO) improves \(\widehat{\bsxi}_n\) over SISO.

Future research will focus on ``live'' joint estimation of \( \V \), \( \boldsymbol{\Sigma}_{\boldsymbol{\epsilon}} \) and \( \boldsymbol{\Sigma}_\w \) without any pre-calibration phase or dedicated measurement (thus allowing for the possibility that these parameters are time-\emph{dependent}) as well as on optimal pilot-tone placement in the available spectrum for best trade-off between jitter estimation and bandwidth usage.

\section{Acknowledgments}

ChatGPT “gpt-5” was used to a limited extent for text harmonization and grammar checks. The authors confirm that all theoretical and numerical contributions, proofs, codes and conclusions are exclusively their own.

\end{document}